# Multi-level resistance switching and random telegraph noise analysis of nitride based memristors

Nikolaos Vasileiadis[1,3*], Panagiotis Loukas[1], Panagiotis Karakolis[1,2], Vassilios Ioannou-Sougleridis[1], Pascal Normand[1], Vasileios Ntinas[3], Iosif-Angelos Fyrigos[3], Ioannis Karafyllidis[3], Georgios Ch. Sirakoulis[3], and Panagiotis Dimitrakis[1*]

[1] *Institute of Nanoscience and Nanotechnology, National Center for Scientific Research "Demokritos", 15341 Ag. Paraskevi, Greece*

[2] *Department of Physics, University of Patras*, 26500 Patras, Greece

[3] *Department of Electrical and Computer Engineering, Democritus University of Thrace, Xanthi, 67100, Greece*

* Corresponding Authors: n.vasiliadis@inn.demokritos.gr, p.dimitrakis@inn.demokritos.gr

*Abstract*—**Resistance switching devices are of special importance because of their application in resistive memories (RRAM) which are promising candidates for replacing current nonvolatile memories and realize storage class memories. These devices exhibit usually memristive properties with many discrete resistance levels and implement artificial synapses. The last years, researchers have demonstrated memristive chips as accelerators in computing, following new in-memory and neuromorphic computational approaches. Many different metal oxides have been used as resistance switching materials in MIM or MIS structures. Understanding of the mechanism and the dynamics of resistance switching is very critical for the modeling and use of memristors in different applications. Here, we demonstrate the bipolar resistance switching of silicon nitride thin films using heavily doped Si and Cu as bottom and top-electrodes, respectively. Analysis of the current-voltage characteristics reveal that under space-charge limited conditions and appropriate current compliance setting, multi-level resistance operation can be achieved. Furthermore, a flexible tuning protocol for multi-level resistance switching was developed applying appropriate SET/RESET pulse sequences. Retention and random telegraph noise measurements performed at different resistance levels. The present results reveal the attractive properties of the examined devices.**

I. INTRODUCTION

Resistive memories (RRAMs) are among the strongest candidates for emerging nonvolatile memories. Owing to their excellent scalability and simple two-terminal structure, RRAMs are ideal for implementation in crossbar architecture, which is the most promising alternative to achieve the smallest memory cell, $4F^2$ (F: the minimum feature size obtained by lithography) [1]. The validity of RRAM crossbar (Xbar) arrays to implement in-memory [2, 3] and neuromorphic computing accelerators [4] has been demonstrated. Recently, we have shown that resistive switching (RS) devices can be used as memristors to store qubits in quantum simulators [5]. From the technology point of view [6], the most attractive materials are the metal oxides (OxRAM) and the phase-change materials (PCRAM). It is well known that OxRAM properties are severely affected by the environmental conditions [7] and impurities [8, 9]. Hence, humidity and oxygen diffusion barriers should be used in order to keep oxide's functionality stable [10]. In this context, silicon nitride-based insulators are of special importance because of their immunity against humidity and oxygen related parasitic effects [11, 12] as well as against metal ion diffusion [13, 14]. Silicon nitride is a well-known material in NVM technology and various charge-trapping memory devices are in the market (e.g. SONOS, BiCS). The reason for this success is the presence of intrinsic bulk defects, acting as trapping levels for both electrons and holes [15,16]. Recently, the use of silicon nitride for RRAM and memristors has been successfully demonstrated [13-15]. Furthermore, the properties of silicon nitride memristors can be tuned by intentional doping incorporation [17]. Reliable neuromorphic computing and the realization neuromorphic chip based on memristors is a significant breakthrough especially for edge-computing applications. Thus, the dynamic operation of memristive devices is very important. For the evaluation of the dynamic operation, the resistance tuning protocol to access multiple resistance levels and their retention should be investigated. Furthermore, the resistance fluctuations of each level should be characterized.

In this contribution, we report on resistive switching in metal-nitride-silicon memristive devices with emphasis on the role of the traps. Section II describes the tested devices and the measurement setups used for their characterization. Stability characterization of the various resistance levels performed through random telegraph noise (RTN) measurements and analysis. Section III presents the experimental results and their analysis. The conclusions are summarized in section IV.

II. DEVICE FABRICATION AND MEASUREMENTS SETUP

On a n$^{++}$-Si wafer ($\varrho < 0.003$ Ω.cm), a 7 nm Si$_3$N$_4$ layer was deposited by LPCVD at 810 °C, using ammonia

(NH$_4$) and dichlorosilane (SiCl$_2$H$_2$) gas precursors (sample S1). On another n$^{++}$-Si wafer, the same nitride layer was deposited above a 2nm thermally grown SiO$_2$ layer (sample S2). Both wafers, top-electrodes were defined by photolithography and metal lift-off. Metallization comprises a sputtered 30nm Cu layer covered by 30nm Pt to prevent oxidation of Cu. The role of the SiO$_2$ layer which was placed in between SiN and Si is mainly twofold. First, to control the tunneling of carriers from n$^{++}$ - Si (BE) to SiN and vice versa. Second, to increase the lifetime (retention) of the achieved resistive states adding a higher energy barrier and subsequently mitigate the leak of trapped carriers from SiN to Si BE. Furthermore, the quality of the interface between the SiO2 (thermal dry oxide) and Si-BE is significantly higher that the SiN/Si one. Therefore, it is expected that the additional SiO$_2$ layer will contribute to the reduction of noise in the electrical measurements for S2 device due to the thermally activated exchange (trapping/de-trapping) of carrier between BE and SiN.

DC I-V measurements were performed at room temperature on a wafer prober using a Keithley 4200 semiconductor parameter analyzer (SPA). Dynamic measurements were also employed in order to investigate the multi-level resistance tuning of the examined memristors and to perform RTN measurements. The memristor was installed on a probe station and triax cables were used to supply the required voltage pulses and read the output currents. For multi-level resistance investigations, a precise control of compliance current and flexible pulse tuning protocol is necessary. For this purpose, we built a custom setup where a DAQ-card NI-PCIO-MIO-16E is connected with a SR570 I/V converter through a low-noise junction box (NI BNC 2110). Figure 1(a) shows the block diagram of the experimental setup used. To achieve a fast and accurate passive compliance current mechanism, a pair of NMOS-PMOS transistors (ALD1116, ALD1117) was utilized. During SET (RESET) operation, the NMOS is connected (disconnected) and the PMOS is disconnected (connected). This selection between NMOS and PMOS is performed by a reed relay (HE3321C0500) while the memristor is kept under zero bias. The same relay approach was followed for I/V connection (disconnection) during READ (SET/RESET), as shown in fig.1(b). The drain of the transistors is connected to the memristor's BE, while the drive voltages for relays and the pulses applied on the TE of the memristor and on the gates of the NMOS-PMOS pair are provided by the DAQ-card. Figure 1(c) shows the experimental setup used for RTN measurements. Specifically, after setting the memristors at a specified resistance level, the RTN was recorded by digital oscilloscope Agilent 7000 series Oscilloscope (DSO), for 100 s with a sampling time 25 μs, via an I/V converter. For all experiments, memristors with an area of 100 μm × 100 μm were used at room temperature.

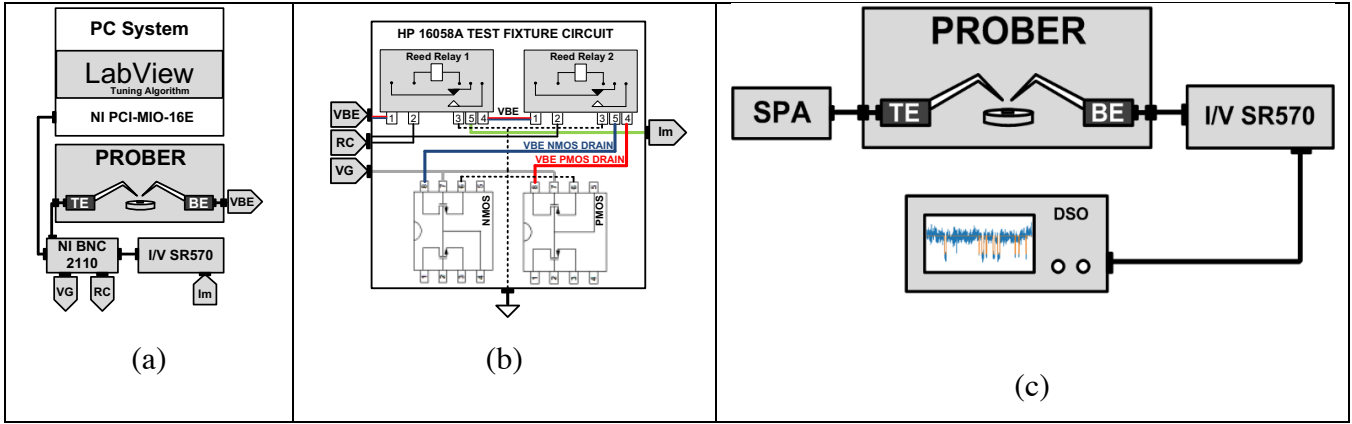

Fig. 1. (a) Block diagram of the used experimental setup, (b) Detail for the connection of external NMOS and PMOS transistors for current compliance, (c) Random telegraph noise measurement setup.

The above setup was controlled by a LabView software allowing the application of any tuning pulsing sequence of arbitrary waveform. A filamentary-change sensitive algorithm was used for precise resistance tuning. Its pseudocode is included in Table 1 below. More specifically, the algorithm constantly searches the appropriate tuning voltage through small increases in applied voltage amplitude. A positive pulse is applied when the current resistance value is above the targeted $R_{max}$ resistance threshold and a negative pulse is applied when the current resistance is below $R_{min}$. When a significant resistance jump $\Delta R$ is observed ($\Delta R > m|R_{max}-R_{min}|$), the tuning voltages return to the initial values, otherwise if a predefined number of tuning attempts are achieved with the no significant resistance jumps ($\Delta R < n|R_{max}-R_{min}|$), then the tuning voltage amplitude increases. In our case, we chose a non-linear voltage increasing method described by the following relation $Ramp(\alpha \cdot log(|\Delta L|/|\Delta R|))$ which takes into account the resistance jump $\Delta R$ as a fraction of change of the overall resistance "length" $\Delta L$ need to be covered for reaching the targeted resistance value ($(R_{max}-R_{min})/2$). Due to logarithmic nature, the above empirical relation was used for faster convergence prolonging the switching endurance of the tested devices against thermal damage [18,19,20]. For avoiding any misinterpretations, the algorithm applies a READ pulse 0.1 V / 200 μs after every tuning pulse, and it stops applying tuning pulses when the targeted resistance range reached ($R_m \in [R_{min}, R_{max}]$). After that only READ pulses are applied until the program stops by the user.

Table 1.

| Flexible Multi-Level Tuning Algorithm Pseudocode |
|---|
| **Initial inputs:** Pulse width, VPOS_INIT, VNEG_INIT, VG, α, n, m, Rmax, Rmin, ATTEMPTMAX |
| 1: **If** Rm ∈ [Rmin, Rmax] |
| 2:     VPOS = VPOS_INIT, VNEG = VNEG_INIT; |
| 3:     «STOP APPLYING TUNING VOLTAGES» |

```
 4: Elseif Rm > Rmax
 5:     ATTEMPTPOS++, ATTEMPTNEG = 0; VNEG = VNEG_INIT;
 6:     If ΔR > m|Rmax-Rmin|
 7:         VPOS = VPOS_INIT;
 8:     Elseif ATTEMPTPOS >= ATTEMPTMAX && ΔR < n|Rmax-Rmin|
 9:         VPOS = VPOS + Ramp(α·log(|ΔL|/|ΔR|));
10:         ATTEMPTPOS = 0;
11:     End
12:     «APPLY TUNING VOLTAGE VPOS»
13: Elseif Rm < Rmin
14:     ATTEMPTNEG ++, ATTEMPTPOS = 0; VPOS = VPOS_INIT;
15:     If ΔR > m|Rmax-Rmin|
16:         VNEG = VNEG_INIT;
17:     Elseif ATTEMPTNEG >= ATTEMPTMAX && ΔR < n|Rmax-Rmin|
18:         VNEG = VNEG - Ramp(α·log(|ΔL|/|ΔR|));
19:         ATTEMPTNEG = 0;
20:     End
21:     «APPLY TUNING VOLTAGE VNEG»
22: End
```

## III. EXPERIMENTAL RESULTS AND DISCUSSION

### A. *DC Current – Voltage sweeps*

Typical I-V sweep measurements are shown in figure 2. The bipolar SET/RESET operation is evident. No electroforming step is required. Also, in the same figure we demonstrate different resistance levels achieved by applying different current compliance ($I_{CC}$) levels. The $I_{CC}$ is aiming to control the size of the low-resistance area (LRS) and to prevent dielectric's failure. In our case, a common current overshoot is observed at LRS during the RESET sweep. This is due both to the fact that the measurement instrument cannot stop the fast-current peak and to the discharge of the device's self-capacitance [21, 22].

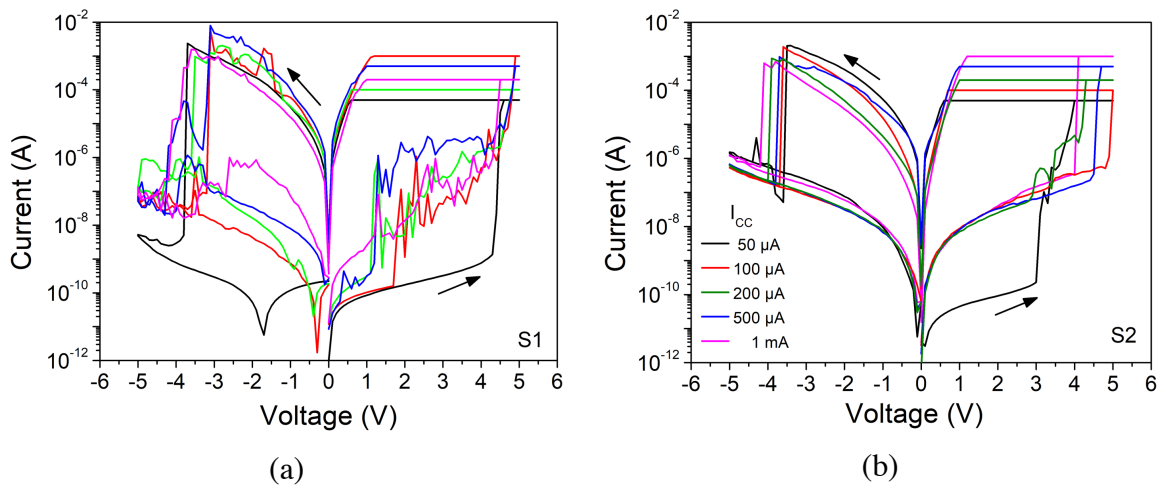

Fig. 2. Typical bipolar switching behavior for (a) S1 and (b) S2. Different resistance levels achieved under different current compliance values.

Evidently, the I-V characteristics for S1 (Fig.2a) exhibits stronger current fluctuations during SET and RESET compared to S2 (Fig. 2b) that is coherent with our initial dielectric stack design considerations. Application of positive bias on the TE of our devices can cause the generation of $Cu^+$ cations, nitrogen vacancies $V_N$ and/or injection of electrons from the $n^{++}$ Si substrate. The fact that the current fluctuations are sufficiently suppressed by the insertion of a thin $SiO_2$ (S2 case) suggests that the mechanism responsible for the current fluctuations in S1 is due to the exchange of electrons between the $n^{++}$ Si substrate and silicon nitride traps. The electrons can be easily injected into short-range traps is promoted by the low energy barrier between Si and $SiN_x$, 5.1-5.3 eV [23, 24] and the large concentration of insulator's intrinsic traps [15,16,25].

The origin of the resistance switching (RS) in silicon nitride is not still well known. Nevertheless, the majority of research results converge to the conclusion that RS originates from a trap-assisted mechanism [26,27], except [28] where the most probable mechanism is attributed to the movement of protons due to the large concentration of hydrogen atoms. This is mainly due to the different deposition techniques that affect the thermodynamic parameters of the defect formation [29]. In our case, the space charge limited conduction (SCLC) mechanism was best fitted to our I-V measurements (Fig.3), and this is the most common mechanism found in the literature for $SiN_x$ memristors [26]. Initially, we have the tunneling of electrons to short-range defects in the nitride layer (Ohm's law) and as the voltage increases more traps deeper in the layer are filled causing the transition, at $V_{TR}$, from linear to parabolic I-V dependence. When all traps are filled, at $V_{TFL}$, the current suddenly increases. According to the SCLC theory the concentration of traps can be estimated from

$$N_t = \frac{2\varepsilon V_{TFL}}{qd^2} \qquad (1)$$

Where $\varepsilon$ is the vacuum dielectric constant of silicon nitride, $q$ is the fundamental electronic charge, while $d$ and $V_{TFL}$ denote the insulator's thickness and the trap-filled limit voltage respectively. According to (1), the trap concentration in $SiN_x$ layers were estimated to $1.14 \times 10^{20}$ cm$^{-2}$ and $1.2 \times 10^{20}$ cm$^{-2}$ for S1 and S2 samples respectively, which are typical values for such materials, suggesting also that the nitride layers have similar properties [15,16,25]. Thus, the differences of the operation parameters between S1 and S2 are mainly attributed to the presence of the oxide layer. Furthermore, Comparing the I-V curves for S1 and S2 in Fig.2, we conclude that reset I-V curves of S1 exhibit transitions to higher current levels (e.g. $I_{CC}$=1 mA). This is due to the thermal recovery from the reset breakdown [30].

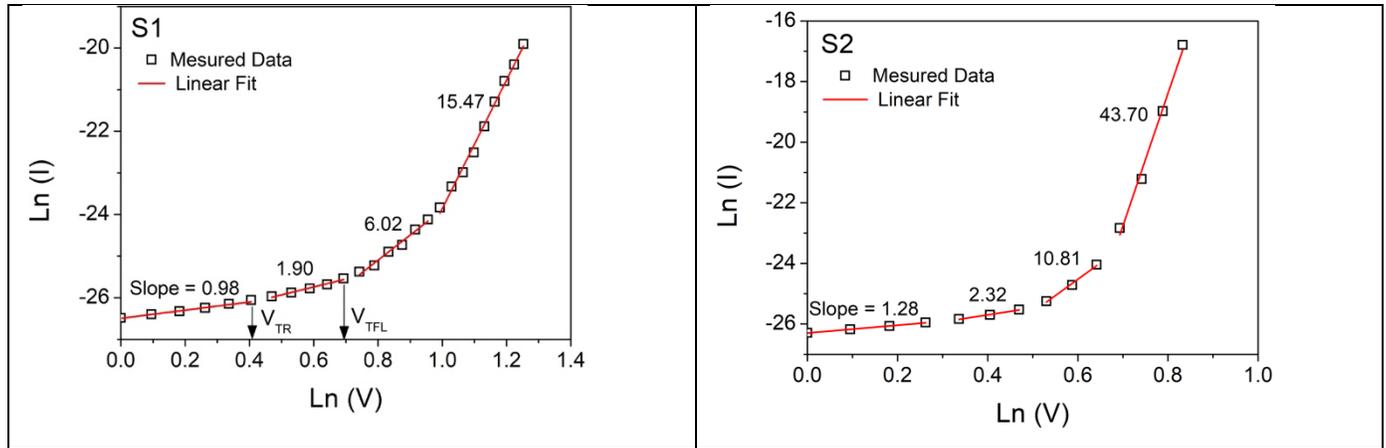

Fig.3 Analysis of typical I-V sweep characteristics following the SCL conduction mechanism for (a) S1 and (S2) samples.

In Fig. 4(a), the statistics for set and reset are presented. Resistance window is slightly higher in S1 compared to S2. The variability in set reset voltages is higher for S2.

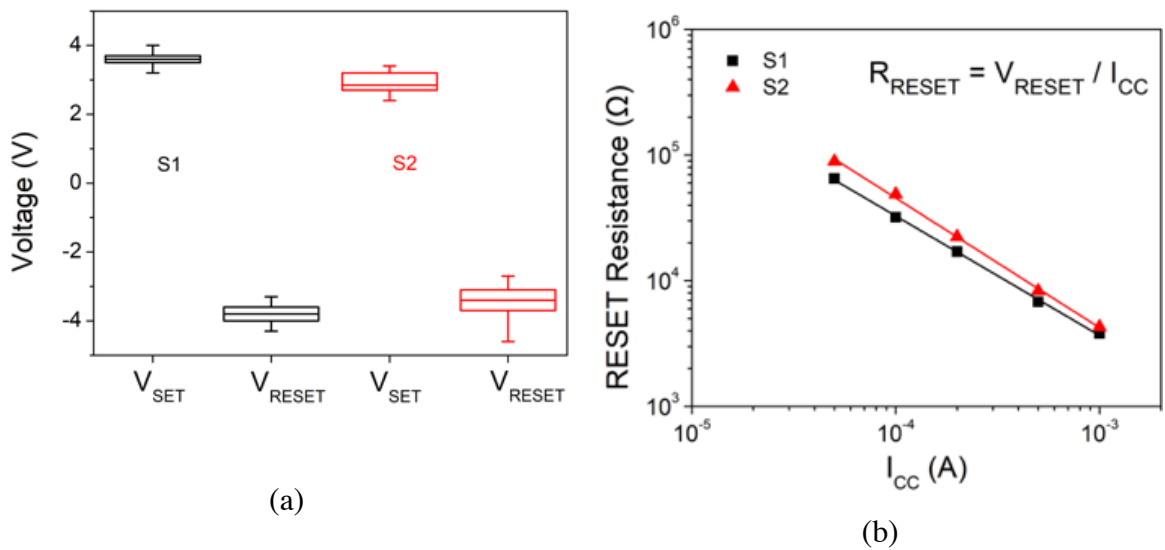

Fig. 4. (a) Statistical box charts for $V_{SET}$ and $V_{RESET}$ for the examined samples. (b) Dependence of RESET resistance on the current compliance for the examined devices.

Figure 4(b) demonstrates the linear behavior between reset resistance and $I_{CC}$ leveraging the results in fig. 4. The slope S1 is (-0.961 ± 0.016) (Ω/A) and for S2 (-1.03 ± 0.02) (Ω/A). According to [31] this is an indication that the RS mechanism is governed by the power provided to our memristors rather than the energy.

### B. *Precise resistance tuning through pulse and compliance modulation*

In order to investigate further the RS mechanism in our devices, without the effect of $I_{CC}$ overshoot, we use the experimental setup presented in Fig.1. The software executed the following algorithm: Starting with the definition of the $R_{min}$ and $R_{max}$ thresholds, the software applies positive (or negative) pulses repeatedly as long as the resistance of the device is above (or below) the targeted thresholds. Specifically, the algorithm stops supply pulses if the resistance is between the targeted thresholds.

Figure 5(a) presents the experimental results after a sequential resistance tuning in the range 200 kΩ → 500 kΩ → 800 kΩ → 500 kΩ → 200 kΩ for sample S2. For this experiment were used the following initial conditions [Pulse width = 10us, $V_{POS\_INIT}$ = 4V, $V_{NEG\_INIT}$ = -4V, $V_G$ = 0.8V, α = 0.1, n = 0.5, m = 5, $R_{max}$ = Rm+50KΩ, $R_{min}$ = Rm-50KΩ, ATTEMPT$_{MAX}$ = 10]. The waiting time at each resistance level is notified by colored straight lines and varies between 20-30s. During 20s, ca.650 READ operations (0.1V/200μs) were performed. In Fig. 5(b), the tuning of the pulse height, until the desired resistance level is achieved, is presented ($V_{TE}$). The compliance current is controlled by the gate voltage ($V_G$) on a MOSFET and remains constant during tuning process. Once the required level is achieved, only READ pulses are applied every 30 ms.

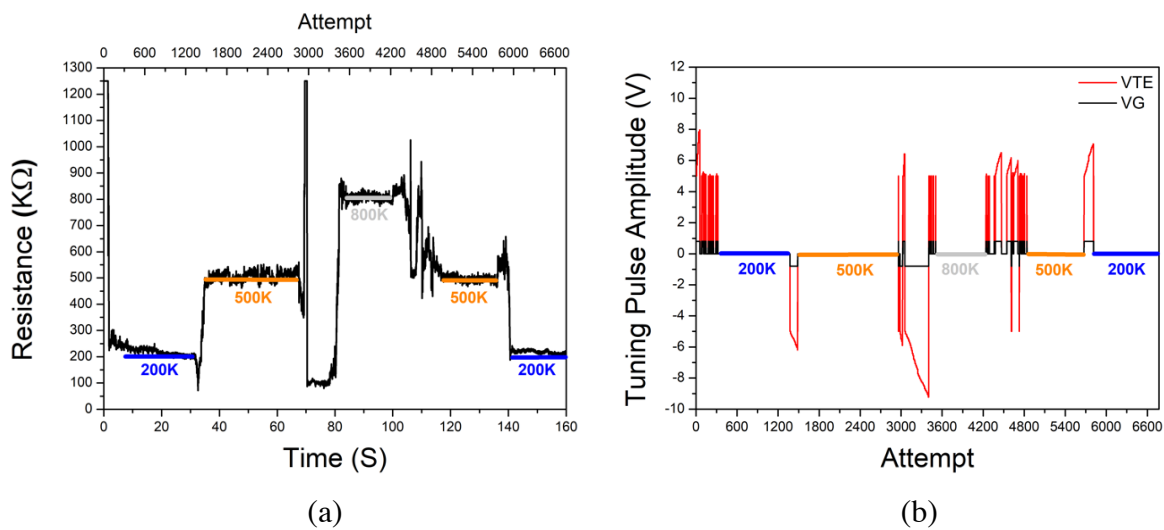

Fig. 5. (a) Sequential resistance tuning with stability periods of 20 seconds on S2 sample, and the corresponding tuning pulse sequence (b).

## C. RTN and multi-level resistance stability investigations

After achieving a stable filament with the above algorithm, we switch to the setup shown in figure 1(c) to take the RTN measurement. A semiconductor parametric analyzer was used as a precision voltage source at 0.1V during the RTN measurement. The sampling time of the RTN data recording was 25μs, which carried out by an Agilent 7000 oscilloscope at high resolution mode with 4Msamples buffer memory size giving 4 MSamples × 25 μs = 100 s long RTN signals. The I/V converter was capable of -6db low pass filtering at 10KHz. In figures 6 (a), (b) twelve read current signals are shown for S1 and S2 respectively. The devices were SET at four dominant resistance levels, 60, 200, 500 and 800 kΩ, and three RTN measurements obtained at each resistance level. The colors were adjusted in the best possible way so that the signals were visible.

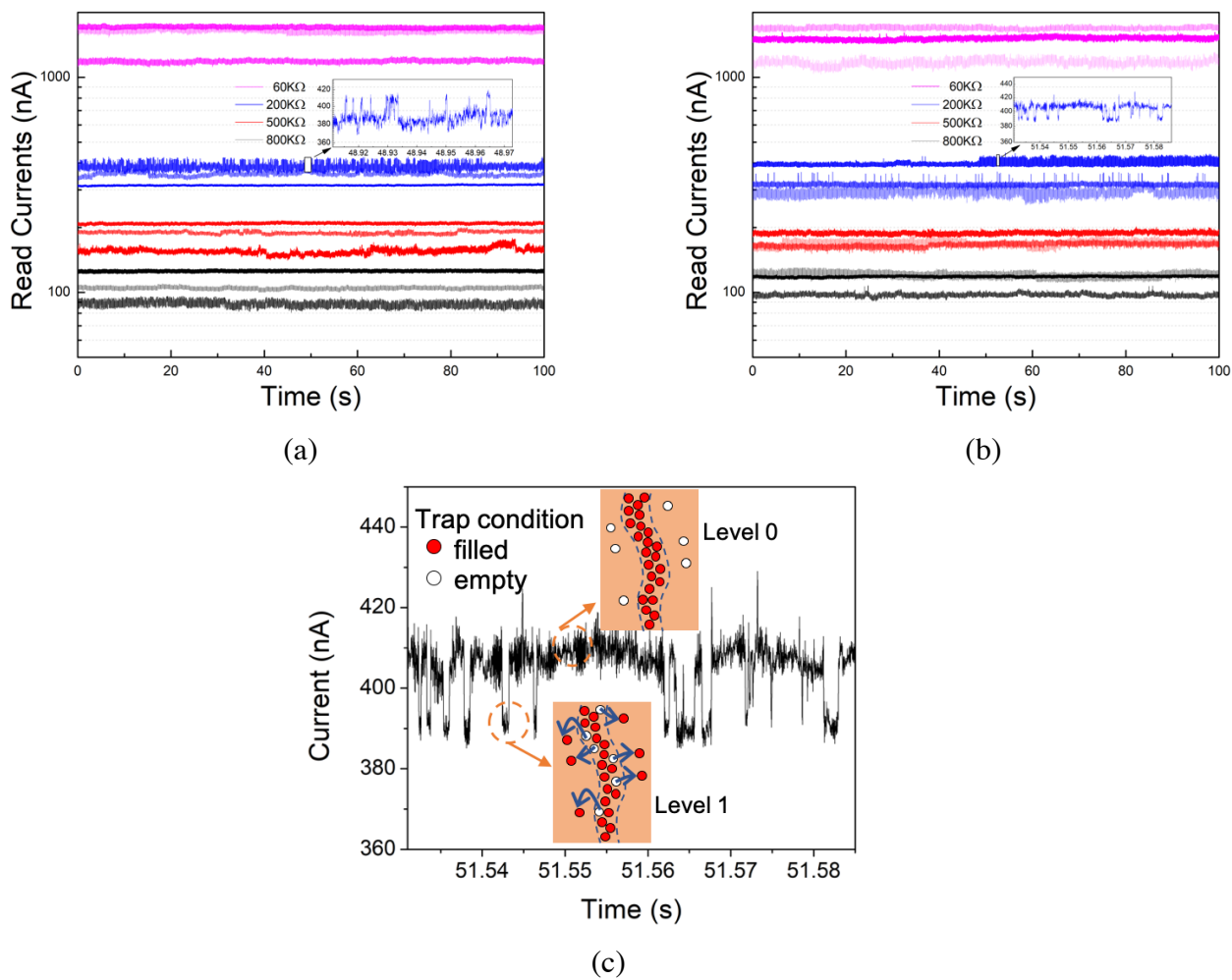

Fig. 6. (a), (b) RTN currents at deferent resistance levels as captured by the oscilloscope for samples (a) S1 (b) S2. (c) Schematic representation of filamentary mechanism explaining the observed RTN signal levels, as measured in sample S2.

Note that it is not observed RTN in all signals, for reasons that will be explained later in this section. The inset plots show typical RTN occurred in each one of the examined samples corresponding to the same resistance level.

The RTN signal in our devices is originated from the escape of electron from the conductive filamentary (CF) area. This is schematically shown in Fig. 6(c), where the current is flowing through the programmed conductive filamentary area corresponding to a specific resistance value, Level 0. Spontaneously, electrons are moving from the CF to the surrounding traps, where they stay for a limited τ and return back to the traps in the CF. This process reduces the read current (i.e. increases the resistance of the resistance of the CF) for a short time leading to negative current bursts.

The RTN signals shown in Fig. 6 were analyzed. For the sake of comparison, we will focus on 200 kΩ level for S1 and S2. Figure 7 summarizes the RTN analysis results of 200 kΩ resistance level for both devices. Obviously, in Fig.7(a) and (d) there exist two discrete levels in RTN data with different time durations. Evidently, the current jumps between the discrete levels are higher in S1. The distribution of the amplitude of the discrete levels in read current is presented in Fig.7(b, e), where clearly reveals a bimodal current (resistance) distribution corresponding to the observed discrete current levels. These metastable levels correspond to different traps and were estimated by fitting a Gaussian function to each distribution. The fitting results are presented in Table 2. Using the mean value of each Gaussian distribution, we reproduce the RTN signal, as shown in Fig.7(a, d).

According to our model for the origin of the observed RTN signal in our devices, we can safely assume that is can be described by the carrier generation-recombination noise model due to a single trapping level, and hence the power spectral density (PSD) of an RTN signal is expressed by the following equation

$$S_I(f) = \frac{\langle \Delta N^2 \rangle}{\langle N \rangle^2} \frac{4/f_T}{1+(f/f_T)^2} \quad (2)$$

Where $\langle \Delta N^2 \rangle$, $\langle N^2 \rangle$ are the mean-square of the carrier fluctuation and the average concentration of the available carriers, f and $f_T$ is the measurement frequency and the release frequency during trapping/de-trapping and $S_I$ is expressed in nA²/Hz. If the RTN signal is due to more than one trapping levels (i.e. traps with different activation energies) then

$$S_I(f) = \sum S_i(f) = \frac{\langle \Delta N^2 \rangle}{\langle N \rangle^2} \frac{4/f_T}{1+(f/f_T)^n} \quad (3)$$

Where $S_i(f)$ is the PSD of i-th trapping level and the exponent n is lying in the range 1≤n<2. McWhorter [32] proved that if we have a large number of different trapping levels then n®1 and hence the G-R noise behaves like 1/f flicker noise making effects from individual trap level noise impossible to resolve. Hooge [33] proved that Eq.(3) is correct when the number of free carriers is larger than the trapped carriers.

Fitting of our experimental PSD data in log-log plots, reveal that two different regions exist in the examined frequency range [0.001 Hz, 400 kHz] in both devices. Specifically, in the low-frequency regime (< 300 Hz) n is very close to 1 and at higher frequency [800 Hz, 8 kHz] is 1.74 and 1.291 for S1 and S2 respectively.

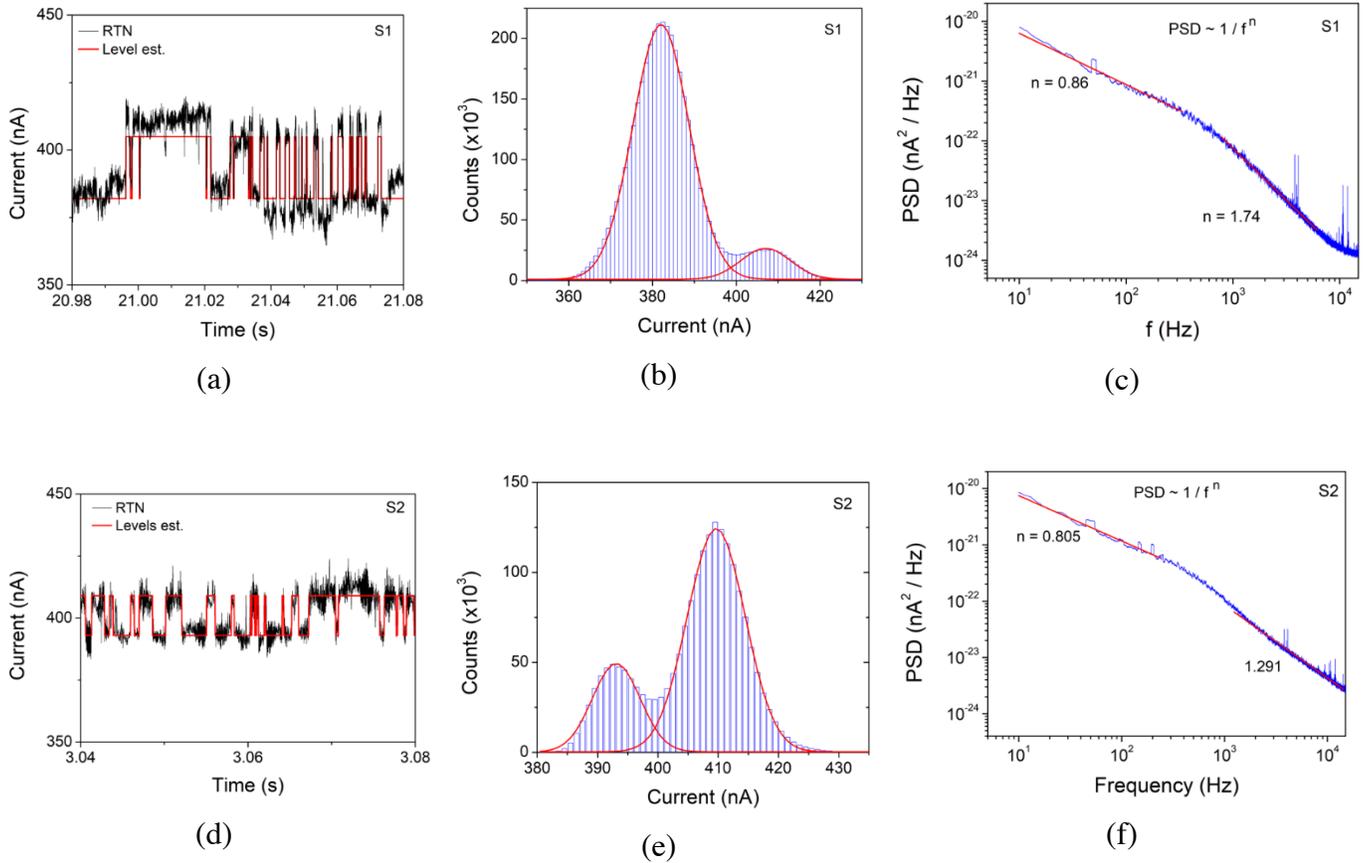

Fig. 7. (a,d) RTN signal and the estimated current discrete levels, (b,e) RTN histogram, (c,f) the related PSD plot at 200 kΩ state of S1 and S2 sample respectively.

Following a methodology similar to [34], the distribution of the dwell times was calculated and is plotted in Fig.8(a, b, d, e). The distribution for each level can be fitted to a Poisson function

$$P(t) = \frac{\Delta t}{\tau} \exp\left(-\frac{t}{\tau}\right) \qquad (2)$$

Where τ is the characteristic time [34]. The characteristic times for each level are shown in Table 2, and are indicatives for the lifetime of an electron in each metastable level. The fitting of our experimental distributions to Poisson suggest that our RTN signal is due to stochastic events such as the trapping and de-trapping of electrons in the vicinity of the CF. Furthermore, the characteristic frequency 1/τ is temperature dependent according to the following Arrhenius relation [34]

$$\frac{1}{\tau} = \frac{1}{\tau_0} \cdot exp(-E_A/k_B T) \qquad (5)$$

Where $1/\tau_0$ could be assumed as the mean thermal vibration frequency of the lattice, i.e. $10^{13}$ Hz [35]. From Eq.(5) using the estimated values of $\tau$ we can estimate the activation energy of the traps corresponding to the two current discrete levels for S1 and S2. The results are shown in Table 2.

Table 2. Summary of the estimated parameters from RTN data analysis

| Device | S1 | S2 |
|---|---|---|
| Gauss. Level 0, mean value (nA) | 382.00 ± 0.03 | 393.00 ± 0.07 |
| Gauss. Level 0, std. (nA) | 7 | 4 |
| Gauss. Level 1, mean value (nA) | 407.00 ± 0.03 | 410.00 ± 0.03 |
| Gauss. Level 1, std. (nA) | 6 | 5 |
| Low-f PSD Slope | 0.858 ± 0.001 | 0.805 ± 0.001 |
| High-f PSD Slope | 1.741 ± 0.001 | 1.291 ± 0.001 |
| Level 0, Dwell time constant $\tau$ (s) | 0.0028 ± 0.0001 | 0.0078 ± 0.0005 |
| Level 1, Dwell time constant $\tau$ (s) | 0.077 ± 0.004 | 0.0042 ± 0.0001 |
| Level 0, Trap Energy (eV) | 0.622 ± 0.001 | 0.648 ± 0.002 |
| Level 1, Trap Energy (eV) | 0.708 ± 0.001 | 0.632 ± 0.001 |

Furthermore, Fig.8(c, f) presents the related weighted time lag plots (wTLP) for each sample. A TLP is constructed by plotting the i-th point of the RTN in the *x*-axis and the (i + 1)-th point in the y-axis for the full RTN trace. Through this signal autocorrelation, it is possible to evaluate whether the values in a time series dataset are random or present concentrations around specific values. These values are related to the various trapping / de-trapping levels [36]. However, if the TLP constructed from RTN signals with background noise and especially from large datasets, these regions are overlapped and the information in them are not well understood. So, in our case we used the weighted TLP method [37], which took into account each point as a small gaussian region and after that adding and normalizing the sum. In figure 8 (c,f) wTLPs reveal the presence of two well separated poles corresponding to the RTN discrete levels for each device. It should be mentioned that the color bar indicates the probability to find a current value at certain time $I(t_{i+1})$ versus the previous value $I(t_i)$. In wTLP, the peaks at the diagonal indicate the RTN levels, while peaks outside the diagonal correspond to transitions between these current levels [37].

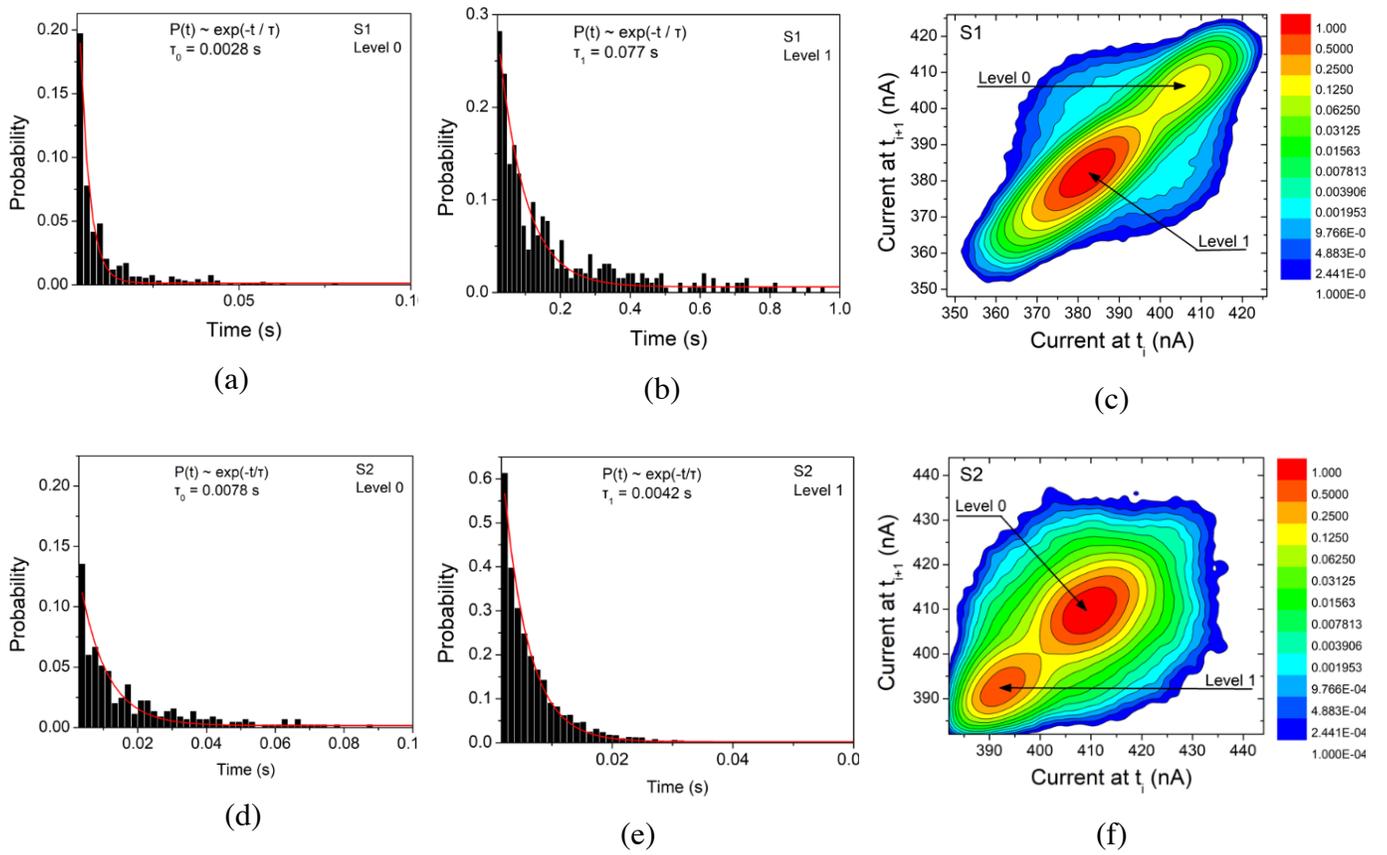

Fig. 8. (a, b, d, e) Histograms of the estimated dwell times for the observed discrete current levels of S1 and S2. (c,f) Weighted time lag plots of the RTN signal at 200 kΩ state of S1 and S2 sample respectively.

## IV. Conclusion

In this work we investigated the multi-level characteristics of SiN$_x$-based memristors. A DC analysis was made reviling the conduction mechanisms of the proposed devices and the analog filamentary capabilities was exploited through the development of an experimental setup and a flexible tuning protocol that allowed accurate multi-level resistance switching. RTN measurements were carried out and an extensive comparison between same resistance level filament, on two asymmetrical by the bottom electrode silicon nitride memristors.

## Acknowledgements

This work was supported in part by the research projects "MEM-Q" (MIS 5021467)" which is implemented under the Action "Reinforcement of the Research and Innovation Infrastructure," funded by the Operational Programme "Competitiveness, Entrepreneurship and Innovation" (NSRF 2014-2020) and co-financed by Greece and the European Union (European Regional Development Fund).